   \documentclass[prl,showpacs,notitlepage,twocolumn]{revtex4}
\usepackage{amsmath}
\usepackage{amssymb}
\usepackage{bm}
\usepackage{epsfig}
\usepackage{graphicx}
\usepackage{color}
\usepackage{natbib}
\usepackage{hyperref}

\newcommand{\Tr}{\mathop{\rm Tr}\nolimits}

\renewcommand{\i}{{\mathrm i}}
\newcommand{\e}{{\mathrm e}}
\renewcommand{\d}{{\mathrm d}}
\newcommand{\av}[1]{\left\langle #1\right\rangle}
\begin{document}
\title{
% Entangled photons emission by quantum dot in microcavity: the role of pumping\\
%   Emission of entangled photons from microcavity with quantum dot\\
% Stationary and pulsed entangled photons emission from quantum microcavity\\
% Semiconductor microcavity as a continuous or pulsed source of entangled photons
Dramatic impact of pumping mechanism on photon entanglement in microcavity
}
\author{Alexander N. Poddubny}

\affiliation{Ioffe Physical-Technical Institute RAS, 26 Polytekhnicheskaya, 194021 St.-Petersburg, Russia}
\begin{abstract}
A theory of entangled photons emission from quantum dot in microcavity under 
continuous and pulsed incoherent  pumping is presented.
It is shown that the time-resolved 
two-photon
correlations  drastically depend on the pumping mechanism: the continuous pumping quenches the polarization entanglement and strongly suppresses photon correlation times. Analytical theory of the effect is presented.
\end{abstract}

\date{\today}

\maketitle
% % % % % % % % % % % % % % % % % % % % % % % 
% \section{Introduction}\label{sec:intro}
Semiconductor quantum dots are a promising source of single photons and entangled photon pairs. Polarization-entangled photons generated during the radiative recombination of the quantum dot biexciton are now in a focus of intensive experimental
 research~\cite{Akopian2006,hudson2007,Stevenson2008, muller2009,Bennett2010,Dousse2010,finley2011,arakawa2011}. 

The state-of-the art approach to increase the rate of  photon pair generation is to place the dot in the specially designed cavity, where the frequencies of the two different photon modes are independently tuned to the biexciton and exciton resonances~\cite{Dousse2010}. Both exciton and biexciton radiative recombinations are then increased, allowing to observe bright two-photon emission~\cite{Dousse2010}.
Here we study theoretically  the quantum emission properties of such  microcavity  under incoherent pumping.  We analyze the polarization density matrix of the photon pair, determined by the second-order correlation function of the photons $g^{(2)}$~\cite{james2001}. 
 The incoherent pumping itself is an intrinsic feature of the light source. However, to the best of our knowledge, the  pumping effect on the biexciton emission has not been theoretically analyzed yet,  despite the extensive amount of the studies done~\cite{Johne2008,Stevenson2008,laussy2010,arakawa2011}. Experimental design of Ref.~\cite{Dousse2010} has not been addressed theoretically as well.

\begin{figure}[tb]
 \begin{center}
  \includegraphics[width=\linewidth]{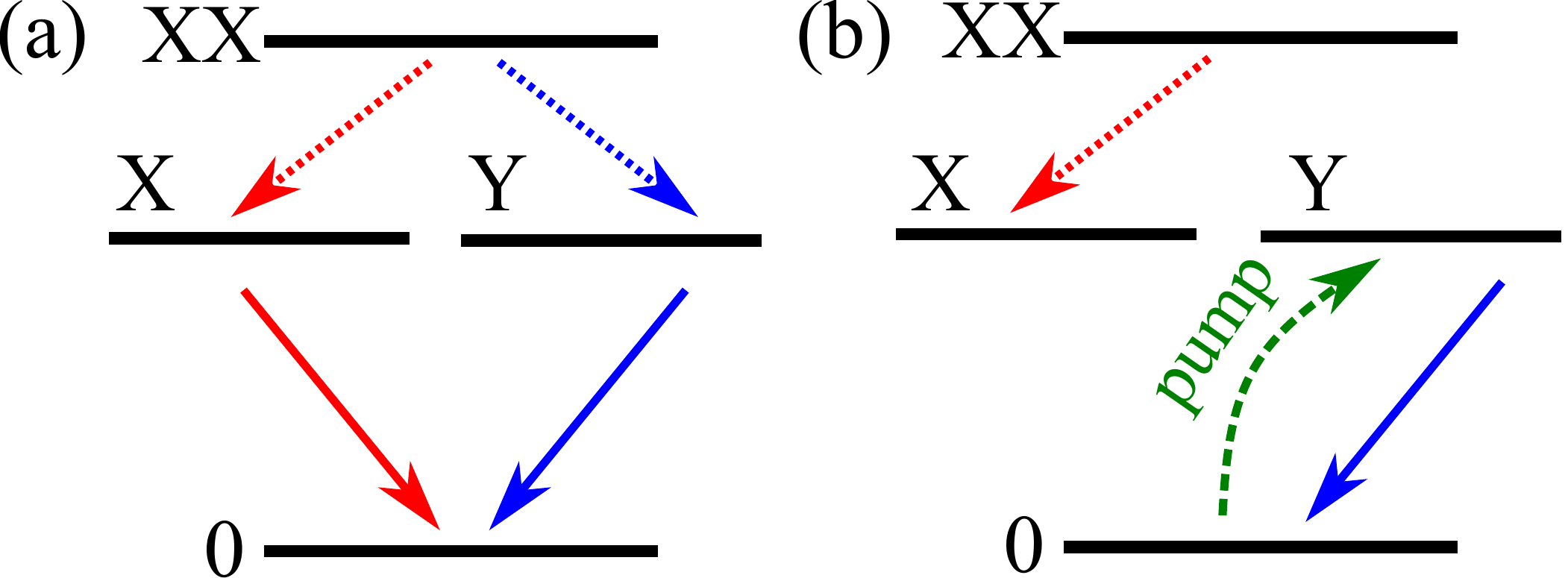}
 \end{center}
\caption{(Color online) (a) Scheme of the two-photon emission from a quantum dot.
Red and blue arrows correspond to $x$- and $y$-polarized photon modes, tuned to the exciton (solid arrows) and biexciton (dotted arrows) resonances.
The letters 0, X,  Y and XX denote ground state , two bright exciton states and biexciton state, respectively. Panel (b) schematically presents the breakdown of the entangled photon emission at continuous pumping.
}\label{fig:Cascade}
\end{figure}

Here we demonstrate that the entanglement is strongly suppressed at incoherent continuous pumping. The qualitative explanation of this effect is presented on Fig.~\ref{fig:Cascade}.
 Fig.~\ref{fig:Cascade}(a) schematically illustrates the cascade of biexciton emission. The radiative recombination of the biexciton leads to the generation of either two horizontally ($x$) polarized (red arrows), or two vertically ($y$) polarized (blue arrows) photons in the cavity. When  anisotropic  exchange splitting of the bright exciton state\cite{Ivchenko2005} vanishes, these two  channels have the same probability, leading to the completely entangled two-photon state. The situation changes dramatically when the excitons are  generated in the quantum dot continuously. One of the possible mechanisms of the entanglement suppression is schematically illustrated on Fig.~\ref{fig:Cascade}(b). At the first step the biexciton emits $x$-polarized photon (dotted arrow). After that one $x$-polarized exciton is remained in the dot. Due to the pumping, another exciton with  $y$ polarization can be generated (curved green arrow). This $y$-polarized exciton can emit photon before the $x$-polarized one, so that  a pair of cross-polarized photons is present in the cavity. Such naive analysis hints that the pumping can suppress the generation of entangled photon pairs and raises the demand for more thorough calculation.

To elaborate on this effect we have performed a rigorous simulation based on the master equation for the density matrix of the system~\cite{Carmichael}.
% % % % % % % % % % % % % % % % % % % % % % % 
% \section{Model}\label{sec:model}
% % % % % % % % % % % % % % % % % % % % % % % 
% % % % % % % % % % % % % % % % % % % % % % % 
We consider a zero-dimensional microcavity with single  quantum dot. 
% Two photonic modes of the microcavity are independently tuned to the exciton and biexciton resonances of the dot, as in Ref.~\onlinecite{Dousse2010}. 
The Hamiltonian $\mathcal H$  can be written in the following  form,
\begin{equation}\label{eq:H}
 \mathcal H=\mathcal H_{\rm exc}+\mathcal H_{\rm phot}+\mathcal H_{\rm exc-phot}\:,
\end{equation}
where the three terms are the Hamiltonians of the dot, of the cavity photons and of their interaction, respectively.
The quantum dot is grown along $z\parallel [001]$ axis from a zinc-blende semiconductor. Only heavy-hole excitons are taken into account.
Under the above assumptions the Hamiltonian of the dot reads~\cite{Ivchenko2005}
\begin{equation}\label{eq:Hx}
  \mathcal H_{\rm exc}=\sum\limits_{\alpha=x,y,d,d'}\hbar\omega_{\rm X}
n_{\rm X,\alpha}+\hbar\omega_{\rm XX} n_{\rm XX}\:.
\end{equation}
The summation in Eq.~\eqref{eq:Hx} is performed over $x$- and $y$-polarized bright heavy-hole exciton states $|X_{x,y}\rangle$\cite{Ivchenko2005} and over two dark exciton states  $|X_{d,d'}\rangle$, with $\hbar\omega_{\rm exc,\alpha}$ being the energies of these states. The singlet biexciton state with the energy $\hbar\omega_{\rm XX}$
is denoted as $|XX\rangle$. The exciton creation operators $a^\dag_{\alpha}$  have the only non-zero matrix elements   $\langle X_{\alpha}|a_{\alpha}^\dag|0\rangle=
\langle XX |a_{\alpha}^\dag|X_{\alpha}\rangle
=1$.
% \end{equation}
Here $|0\rangle $ is the ground state of the system with no excitations in  conduction and valence bands and no photons in the cavity.
The operators of exciton  and biexciton numbers
$n_{\rm X,\alpha}$ and  $n_{\rm XX}$  have the only non-zero matrix elements $
  \langle X_{\alpha}|n_{\rm X,\alpha}|X_{\alpha}\rangle=1$ and 
$\langle XX |n_{\rm XX}|XX\rangle=1$, respectively.
The anisotropic exchange splitting of the bright exciton doublet is ignored for simplicity.
The photon Hamiltonian reads
\begin{equation}
 \mathcal H_{\rm phot}=\sum\limits_{\alpha=x,y}
\hbar \omega_{\rm X} c_{1,\alpha}^\dag
c_{1,\alpha}^{\vphantom{\dag}}+\sum\limits_{\alpha=x,y}
\hbar (\omega_{XX}-\omega_{\rm X})c_{2,\alpha}^\dag
c_{2,\alpha}^{\vphantom{\dag}}\:,
\end{equation}
where $c_{1,\alpha}^\dag$ and $c_{2,\alpha}^\dag$ are the  creation operators for the two cavity modes with given linear polarization $\alpha=x,y$. We  neglect the polarization splitting of the  modes and assume, that they are independently tuned to the exciton and  biexciton resonances, hereafter  we  term modes $1$ and $2$ as exciton and biexciton photon modes, respectively.  This corresponds to experimental situation of   Ref.~\cite{Dousse2010}
% Such geometry with separate tuning of photonic modes corresponds to the experiment of. 
Finally, the light-exciton interaction Hamiltonian reads
\begin{equation}\label{eq:Hint}
 \mathcal H_{\rm exc-phot}=
\hbar g\sum\limits_{\substack{\alpha=x,y,\\\nu=1,2}}
(c_{\nu,\alpha}^\dag a_{\alpha}^{\vphantom{\dag}}+c_{\nu,\alpha}^{\vphantom{\dag}} a_{\alpha}^\dag)
\:,
\end{equation}
where the interaction constant $g$ is chosen
for simplicity real and the same for both modes. 
% We note, that although the exciton photon mode is detuned for the biexciton resonance (and vice versa) by the biexciton binding energy $\omega_{\rm XX}-2\omega_{\rm X}$, which is large as compared to the coupling constant $g$, their interaction is still taken into account in Eq.~\eqref{eq:Hint}.

The incoherent pumping leads to the generation of the excitons in the dot~\cite{JETP2009}.
The excitons can decay both radiatively and nonradiatively.
The photons, created by exciton recombination, leave the cavity by  tunneling through its mirrors, and are detected in experiment.
To account for all these processes, one has to  solve master equation for the density matrix of the system $\rho$:\cite{Carmichael,laussy2010}
\begin{align}\label{eq:rho}
 \frac{\d\rho}{\d t}\equiv &\mathcal L\rho=\frac{\i}{\hbar} [\rho,\mathcal H]
\\\nonumber
&+
\frac{P_{\rm X}}{2}\sum\limits_{\alpha=x,y,d,d'}(2a_{\alpha}^\dag\rho a_{\alpha}^{\vphantom{\dag}}-a_{\alpha}^{\vphantom{\dag}} a_{\alpha}^\dag\rho-\rho a_{\alpha}^{\vphantom{\dag}} a_{\alpha}^\dag )
\\\nonumber
&+
\frac{\Gamma_{\rm C}}{2}\sum\limits_{\substack{\alpha=x,y\\\nu=1,2}}(2c_{\nu,\alpha}^{\vphantom{\dag}}\rho c_{\nu,\alpha}^\dag-c_{\nu,\alpha}^\dag c_{\nu,\alpha}^{\vphantom{\dag}}\rho-\rho c_{\nu,\alpha}^\dag c_{\nu,\alpha}^{\vphantom{\dag}} )
\\\nonumber&+\frac{\Gamma_{\rm X}}{2}\sum\limits_{\alpha=x,y,d,d'}(2a_{\alpha}^{\vphantom{\dag}}\rho a_{\alpha}^\dag-a_{\alpha}^\dag a_{\alpha}^{\vphantom{\dag}}\rho-\rho a_{\alpha}^\dag a_{\alpha^{\vphantom{\dag}}} )
\:.
\end{align}
Here the quantities $P_{\rm X}$, $\Gamma_{\rm X}$ are the exciton pumping and nonradiative decay rates, 
and $\Gamma_{\rm C}$  is the photon decay rate.
 For simplicity  the polarization and spin dependence of these three processes is disregarded.
We note, that the Lindblad terms in \eqref{eq:rho}
describe the generation and decay for both exciton and biexciton states.
%  It is assumed in Eq.~\eqref{eq:rho},  that the pumping and recombination are spin-independent.

In this paper we restrict ourselves to the weak coupling regime for both exciton and biexciton resonances, i.e. $g\ll\Gamma_{\rm C}$. We also note, that in typical experiments $\Gamma_{\rm X}\ll\Gamma_{\rm C}$~\cite{Khitrova2004,Reithmaier2004,Peter2005}.
Our goal is to determine  the two-photon density matrix
\begin{multline}\label{eq:rho2}
 \rho^{(2)}_{\alpha,\beta;\alpha'\beta'}(t,\tau)=\mathcal N\left\langle \chi_{\alpha'\beta'}^\dag\chi_{\alpha\beta}^{\vphantom{\dag}}
\right\rangle\:,\\
\chi_{\alpha\beta}(t,\tau)=c_{1,\alpha}(t+\tau)c_{2,\beta}(t)\:,
\end{multline}
describing the correlations between the biexciton photon emitted at the time $t$ and the exciton photon emitted at the time $t+\tau$.
The angular brackets denote both statistical and quantum mechanical averaging, the  constant $\mathcal N$ in Eq.~\eqref{eq:rho2} is determined from the normalization condition
$\Tr \rho^{(2)}=1$.

Depending on the experimental conditions, two qualitatively different situations can be realized: pulsed pumping and continuous
pumping. The  procedure to determine $\rho^{(2)}$  is different in these two cases.

% % % % % % % % % % % % % % % % % % % % % % % % % % % 
(i) {\it Pulsed pumping}. In this regime we assume that  short single pumping pulse creates the population of excitons and biexcitons in the quantum dot.  After the pump switches off, the excitons start to recombine radiatively. 
% Such formulation corresponds to the spontaneous emission problem, considered for a semiconductor microcavity e.g. in Ref.~\onlinecite{laussy2009bos}.
Assuming that the pulse duration ($5$~ps in experiment of Ref.~\onlinecite{Dousse2010}) is
longer then the typical energy relaxation times of the carriers (being on  subpicosecond scale\cite{Delerue2004}),
but still shorter than then excitonic radiative lifetime $1/\Gamma_{\rm rad} =\Gamma_{\rm C}/(4g^2)$, being on the order of $100~$ps \cite{Dousse2010},  we can separate the calculation into two steps.
First, we find the density matrix $\rho_0$, generated by the pump pulse, from the equation
% \begin{equation}\label{eq:rho0}
 $\mathcal L_{0}\rho_0=0$\:,
% \end{equation}
where the Liouvillian $\mathcal L_{0}$ differs from
the Liouvillian $\mathcal L$ in Eq.~\eqref{eq:rho} by neglecting the coupling term, $g=0$.
Second, we consider the spontaneous decay after the pump is switched off. This process is described by another Liouvillian, $\mathcal L_{\rm decay}$, where 
exciton-photon coupling is retained but the pumping $ P_{\rm X}$ is set to zero. Thus, we obtain a spontaneous emission problem,
with initial conditions determined by the pump\:.
The two-time correlator $\rho^{(2)}(t,\tau)$ then formally reads \cite{Carmichael}
\begin{multline}\label{eq:rho2expl}
 \rho^{(2)}_{\alpha,\beta;\alpha'\beta'}(t,\tau)=\\\mathcal N
\Tr\left[c_{2\beta'}^\dag
 c_{2\beta}^{\vphantom{\dag}}\e^{\mathcal L_{\rm decay}\tau}\left(c_{1\alpha}^{\vphantom{\dag}}\exp^{\mathcal L_{\rm decay}t}\rho_0
c_{1\alpha'}^\dag\right)
\right]\:.
\end{multline}

(ii) {\it Continuous pumping}. % \subsection{Continuous pumping}\label{sec:model_cont}
 In this case it assumed, that  the excitons are continuously generated in the quantum dot. The balance between the exciton generation and decay leads to the formation of the  stationary density matrix $\rho$,  found from the stationary solution of Eq.~\eqref{eq:rho}.
 This density matrix allows us to determine stationary particle numbers.
 The two-photon density matrix \eqref{eq:rho2} depends only on the delay $\tau$ and is given, similarly to Eq.~\eqref{eq:rho2expl}, by
 \begin{equation}\label{eq:rho2expl2}
  \rho^{(2)}_{\alpha,\beta;\alpha'\beta'}(\tau)=\mathcal N
\Tr\left[c_{2\beta'}^\dag
 c_{2\beta}^{\vphantom{\dag}}\e^{\mathcal L\tau}\left(c_{1\alpha}^{\vphantom{\dag}}\rho
c_{1\alpha'}^\dag\right)
\right]\:.
\end{equation}

% % % % % % % % % % % % % % % % % % % % % % % % 
% % Figure for pulsed pumping,  av. over B

\begin{figure}[tb]
 \begin{center}
    \includegraphics[width=\linewidth]{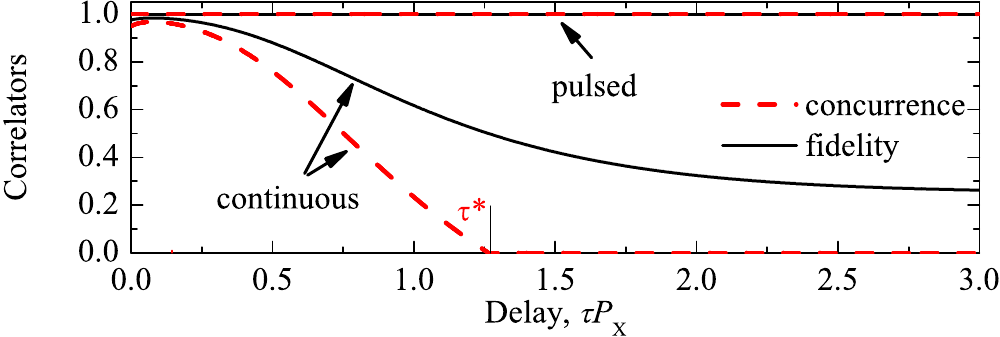}
 \end{center}
\caption{(Color online) Time dependence of concurrence (dashed lines) and fidelity (solid lines)  at continuous and pulsed pumping. Calculation was performed at the following set of parameters: $\hbar g=30~\mu$eV, $\hbar \Gamma_{\rm X}=15~\mu$eV, $\hbar \Gamma_{\rm C}=300~\mu$eV, $\hbar\omega_{\rm XX}-2\hbar\omega_{\rm X}=-5~$meV, $P_{\rm X}=0.1\Gamma_{\rm C}$.
}\label{fig:2}
\end{figure}

% % % % % % % % % % % % % % % % % % % % % % % 
\begin{figure}[tb]
 \begin{center}
    \includegraphics[width=\linewidth]{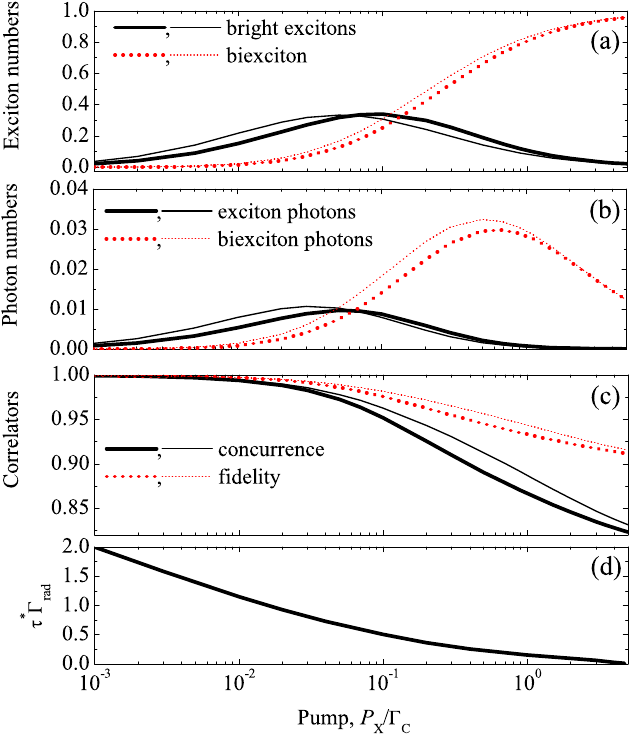}
 \end{center}
\caption{(Color online) Stationary particle numbers and correlators as  function of pumping.
(a) The total numbers of bright excitons (solid curves) and biexciton (dotted curves) photons as  functions  of pumping.
(b) The numbers of exciton (solid curves) and biexciton (dashed curves) photons as  functions  of pumping.
(c) Concurrence (solid curves) and fidelity (dotted curves) of the two-photon pair at zero delay, $\tau=0$.
(d) Concurrence lifetime $\tau^*$, divided by exciton radiative lifetime $1/\Gamma_{\rm rad}$.
Thick curves  are results of numerical calculation, thin curves present the analytical results of Supplemental Materials.
Calculated at the same parameters  as  Fig.~\ref{fig:2}.
}\label{fig:3}
\end{figure}
% % % % % % % % % % % % % % % % % % % % % % % % % % % % % % % % % % % 
The density matrix equations were solved numerically expanding the density matrix over the complete basis of the quantum dot and the cavity  states.
The resulting matrices of Liouvillians are sparse, which makes the   calculation of the matrix exponentials feasible.
Some analytical results in stationary regime will be presented below.

% In the case of the pulsed pumping the photons pair is described by the the Bell state $(|x_1 x_2\rangle +|y_1 y_2\rangle)/\sqrt{2}$, where the subscripts $1,2$ denote exciton and biexciton photons. For such state both concurrence and fidelity are equal to one. 
The obtained two-photon polarization density matrix has the following general structure:
\begin{equation}\label{eq:rho2stac}
 \rho^{(2)}(\tau)=\frac1{2(N_{\parallel}+N_{\perp})}
\begin{pmatrix}
 N_{\parallel}&0&0&N_{\circlearrowright}\\
0&N_{\perp}&0&0\\
0&0&N_{\perp}&0\\
N_{\circlearrowright}&0&0&N_{\parallel}
\end{pmatrix}\:,
\end{equation}
where the order of matrix elements is $x_1x_2$, $x_1y_2$,  
$y_1x_2$, $y_1y_2$. Here the quantities  $N_{\parallel}$ and $N_{\perp}$ determine the probability of the  detection of excitonic and biexcitonic photons with matching and different linear polarizations, respectively. The  matrix element $N_{\circlearrowright}$ describes the correlations of clockwise and counter-clockwise circularly polarized photons.
The structure of the density matrix \eqref{eq:rho2stac} is similar to that obtained in Ref.~\onlinecite{hudson2007}. However, in our case  the  suppression of the entanglement is solely related to the pumping and the quantity $N_{\circlearrowright}$ is real.
The concurrence of the state \eqref{eq:rho2stac}, quantifying the entanglement~\cite{james2001}, reads
\begin{equation}\label{eq:C}
 C=\max\left\{\frac{N_{\circlearrowright}-N_\perp}{N_\parallel+N_\perp},0\right\}\:.
\end{equation}

The results of calculation are presented on Fig.~\ref{fig:2}.
For pulsed pumping one has $N_{\circlearrowright}=N_{\parallel}=1$ and $N_{\perp}=0$, which corresponds to the completely entangled Bell state,\cite{james2001} where fidelity and concurrence are both equal to unity. For continuous pumping both concurrence and fidelity are smaller then unity, and are suppressed at large delays $\tau$. After a certain delay $\tau^*$ the concurrence is zero, i.e. the state is not entangled. Entanglement suppression is directly related to the incoherent nature of the pumping. Indeed, excitons are constantly generated in the dot, and then eventually emit photons. As soon as the delay becomes larger, than the dot repopulation time, the  temporal coherence of the photons is lost. At very large delay one has $N_{\parallel}=N_{\perp}=1/4$ and $N_{\circlearrowright}=0$, which corresponds to completely independent photons.

Fig.~\ref{fig:3} presents more detailed analysis of stationary correlators.
Interestingly,  even the stationary entanglement, calculated at $\tau=0$, is suppressed by the pumping. 
First two panels present the calculated bright exciton,
  biexciton and photon occupation numbers as functions of pumping. Thick lines correspond to the results of the numerical calculation. Thin lines present the analytical results  in weak coupling regime, obtained in Supplemental Materials. Here we briefly summarize them. Neglecting the exciton-photon coupling one can get 
stationary exciton population numbers
\begin{equation}\label{eq:Nx}
    N_{\rm X,\alpha}
=\frac{P_{\rm X}}{\Gamma_{\rm X}+4P_{\rm X}+P_{\rm X}^2/\Gamma_{\rm X}},\quad
 N_{\rm XX}=\frac{P}{\Gamma_{\rm X}}N_{X,x},
 \end{equation}
where $\alpha=x,y,d,d'$\:. 
Eq.~\eqref{eq:Nx} indicates that
at small pumpings $N_{\rm X}$ increases linearly with pumping, and $N_{\rm XX}$ increases quadratically. At high pumping the dot is  in biexciton state, i.e. $N_{\rm XX}\to 1$ and $N_{\rm X}\to 0$.
The photon numbers $N_{\rm C,\alpha}^{(1,2)}$ in the weak coupling regime are given by
\begin{equation}\label{eq:Nc}
 \Gamma_{\rm C}N_{\rm C,\alpha}^{(1)}=2\pi g^2N_{X,\alpha}\varrho_1,\quad
\Gamma_{\rm C}N_{\rm C,\alpha}^{(2)}=2\pi g^2N_{\rm XX}\varrho_2
\end{equation}
where
\begin{equation*}\label{eq:Nrho}
 \varrho_1=\frac{2}{\pi (\Gamma_{\rm C}+\Gamma_{\rm X}+5P_{\rm X})},\quad
\varrho_2=\frac{2}{\pi (\Gamma_{\rm C}+5\Gamma_{\rm X}+P_{\rm X})}\:,
\end{equation*}
and $\alpha=x,y$\:. The right hand sides in Eqs.~\eqref{eq:Nc} 
are the rates of photon generation due to exciton recombination. They are presented in the form similar to the Fermi Golden rule and have simple physical meaning:
the rate is proportional to the population of the correspondent state of the dot, to the square of the interaction constant $g$ and to the effective density of states $\varrho$. The latter is quenched at high pumping due to the fermionic nature of the electrons, which typically leads to the pumping-induced linewidth broadening~\cite{laussy2009,Poddubny2010prb}. Thus, at small pumping the exciton and biexciton photon numbers are proportional  to the population of the corresponding quantum dot states, and at high pumpings they are suppressed by the pumping-induced dephasing.
This explains the behavior of the curves on Fig. \ref{fig:3}b.

At the vanishing pumping one has $N_{\circlearrowright}(\tau=0)=N_{\parallel}(\tau=0)=1$ and $N_{\perp}(\tau=0)=0$ in Eq.~\eqref{eq:rho2stac}, which corresponds to the completely entangled Bell state with concurrence $C=1$, see Eq.~\eqref{eq:C}. The pumping leads to the growth of $N_{\perp}$ and to the suppression of $N_{\circlearrowright}$. This quenches the entanglement, in agreement with Fig.~\ref{fig:3}(c). 
Fig.~\ref{fig:3}(d) demonstrates that the concurrence lifetime $\tau^*$, i.e. the time, during which the state remains entangled,  decreases with pumping, because the dot is faster repopulated.
We note, that the high sensitivity of two-photon correlations to the pumping is a rather general effect, known, for instance, in superradiant emission of the cavities with several resonant quantum dots~\cite{Temnov2009} or for single-dot lasers~\cite{jahnke2009}. The main result of Fig.~\ref{fig:3} is that  the entanglement degree is substantially quenched when the pumping is large enough to make the numbers of exciton and biexciton photons  comparable.

To summarize, we have put forward a general theory of the entangled photon generation from the zero-dimensional microcavity with embedded quantum dot under incoherent pumping in the weak coupling regime.
We have demonstrated that the time-dependent polarization density matrix of the entangled photon pair is very sensitive to the mechanism of pumping, i.e. pulsed one or continuous one.   Analytical theory of this effect has been presented. Important goal of the future studies is to analyze the strong coupling regime,  up to now realized only for the exciton resonance~\cite{Khitrova2004,Reithmaier2004}.

%  One of the important goals of the future studies is to analyze the strong coupling regime of exciton and biexciton resonances with photons, which is up to now realized only for the exciton resonance.\cite{Khitrova2004,Reithmaier2004}

%  To summarize, the time-dependent

\acknowledgements
The author  acknowledges encouraging discussions with M.M. Glazov, E.L. Ivchenko and P. Senellart.
This work has been supported by RFBR, ``Dynasty'' Foundation-ICFPM, and the projects ``POLAPHEN'' and ``Spin-Optronics''.
% % % % % % % % % % % % % % % % % % % % % % % % % % % % % % % % % % % 

% \bibliography{all_cp1251b}

\section*{Supplemental material}
\setcounter{equation}{0}
\renewcommand{\theequation}{S\arabic{equation}}
\section*{S1. Stationary exciton and photon numbers}
% \label{sec:appendix}
Now we obtain the stationary occupation numbers of the quantum dot states and the stationary photon numbers, Eq.~\eqref{eq:Nc} and Eq.~\eqref{eq:Nx}.
We  consider the weak coupling  regime, and assume that the additional condition 
\begin{equation}\label{eq:weak}
 \Gamma_{\rm C}\gg \Gamma_{\rm rad}, \text{ i.e. } \Gamma_{\rm C}\Gamma_{\rm X}\gg g^2,
\end{equation}
 holds.
Our goal is to expand the density matrix in powers of the coupling constant $g$,
\begin{equation}\label{eq:series}
 \rho=\rho_0+\rho_1+\ldots,\quad \rho_j\propto g^j\:.
\end{equation}
 The lowest order contribution to exciton number is determined by $\rho_0$ and does not depend on $g$, while the photon numbers are proportional to $g^2$. 
To perform the expansion we express the total Liouvillian $\mathcal L$ of the system \eqref{eq:rho} as $\mathcal L_0+\mathcal L_1$, where
$\mathcal L_0$ is the Liouvillian neglecting exciton-photon interaction, and
\begin{equation}
 \mathcal L_1\rho\equiv \frac{\i}{\hbar} [\rho,\mathcal H_{\rm exc-phot}]\:.
\end{equation}
The zero-order contribution to the  stationary density matrix is found from the equation
\begin{equation}\label{eq:rho0L0}
 \mathcal L_0\rho_0=0\:.
\end{equation}
Each following term of the expansion \eqref{eq:series} is determined by the recurrence relation
\begin{equation}\label{eq:next}
 -\mathcal L_0\rho_j=\frac{\i}{\hbar} [\rho_{j-1},\mathcal H_{\rm exc-phot}]\:.
\end{equation}
Let us first solve Eq.~\eqref{eq:rho0L0}. Obviously, the matrix $\rho_0$ is diagonal, with the only non-zero matrix elements
\begin{align}\label{eq:form_rho0}
 \langle X,\alpha |&\rho_0| X,\alpha\rangle=N_{{\rm X},\alpha}\equiv N_{\rm X}\:,\\
\langle XX |&\rho_0| XX\rangle=N_{\rm XX}\:,\nonumber\\\nonumber
\langle 0 |&\rho_0| 0\rangle=1-N_{\rm XX}-4N_{\rm X}\:,
\end{align}
where we took into account the normalization condition $\Tr \rho_0=1$.
Substituting the density matrix  in the form \eqref{eq:form_rho0} into  Eq.~\eqref{eq:rho0L0} we obtain the following system of linear equations:
\begin{align}\label{eq:syst0}
&-(\Gamma_{X}+P_{\rm X})N_{\rm X}+\Gamma_{X}N_{\rm XX}+
P_{\rm X}(1-N_{\rm XX}-4N_{\rm X})=0\:\nonumber,\\
&-\Gamma_{X}N_{\rm XX}+P_{\rm XX}N_{\rm X} =0\:.
\end{align}
Solution of this system yields Eq.~\eqref{eq:Nx}.

Now we proceed to the calculation of the stationary photon numbers. We consider the case of the $x$-polarized biexciton photons as an example.
Calculating the commutator Eq.~\eqref{eq:next}
for the term $N_{XX}| XX\rangle \langle XX|$, entering $\rho_0$, we get
\begin{equation}
 \frac{\i}{\hbar} [\rho_{0},\mathcal H_{\rm exc-phot}]=
\frac{\i gN_{\rm XX}}{\hbar} b^\dag |0\rangle \langle 0|a c_{2,x}+h.c.
\end{equation}
Assuming, that the binding energy of the biexciton $\omega_{XX}-2\omega_{\rm X}$ is much larger than $P_{\rm X}$, $\Gamma_{\rm C}$, $\Gamma_{\rm X}$ and $g$, we write down Eq.~\eqref{eq:next}
to find biexciton-related term in $\rho_1$:
\begin{equation}\label{eq:interm}
 \frac{1}{2}(\Gamma_{\rm C}+5\Gamma_{\rm X}+P_{\rm X})\rho_1=\frac{\i g N_{\rm XX}}{\hbar} b^\dag |0\rangle \langle 0|a c_{2,x}+h.c.
\end{equation}
Finding $\rho_1$ and calculating the commutator \eqref{eq:next} again, we obtain the generation rate of the biexciton photons, standing in the r.h.s. of Eq.~\eqref{eq:Nc}. 
We see, that Eq.~\eqref{eq:interm} yields the density of states $\varrho_2$. The value of  $\varrho_2$ is inversely proportional to the decay rate in l.h.s. of Eq.~\eqref{eq:interm}. 
Eq.~\eqref{eq:Nc} is in fact the particular case of Eq.~\eqref{eq:next} to find the matrix $\rho_2$, averaged over the states of the dot.
 Analogous procedure yields the generation rate of the exciton photons. The consideration  differs only in the decay rate of the intermediate state of type $a^\dag |0\rangle \langle 0|c_1$, equal to $\pi/\varrho_1$.

% % % % % % % % % % % % % % % % % % % % % % % % 
\section*{S2. Stationary two-photon density matrix}
% \label{sec:appendix2}
In this Section we present the explicit results for the two-photon stationary density matrix \eqref{eq:rho2stac}. The calculation procedure is generally the same as
that presented above to obtain the stationary photon numbers. However, it requires expansion of the stationary density matrix up to the fourth term in the series over $g$ \eqref{eq:series}, proportional to $g^4$. The calculation is therefore much more tedious, but still feasible.  Thus, we will only present the results:
\begin{align}\label{eq:Npar}
 N_{\parallel}\equiv &\av{c_{2,x}^\dag c_{1,x}^\dag c_{1,x}^{\vphantom{\dag}}c_{2,x}^{\vphantom{\dag}}}=\\&\hspace{1cm}\frac{\pi g^2}{\Gamma_{\rm C}}\left[AN_{1,\rm XX}+
A'N_{2,\parallel}+BN_{\rm XX}
\right]\:,\nonumber
\\\
 \label{eq:Ncirc}
  N_{\circlearrowright}\equiv &\av{c_{2,x}^\dag c_{1,x}^\dag c_{1,y}^{\vphantom{\dag}}c_{2,y}^{\vphantom{\dag}}}=\frac{\pi g^2}{\Gamma_{\rm C}}\left[A'N_{2,\circlearrowright}+BN_{\rm XX}
 \right]\:,\\
 \label{eq:Nperp}
  N_{\perp}\equiv &\av{c_{2,x}^\dag c_{1,y}^\dag c_{1,y}^{\vphantom{\dag}}c_{2,x}^{\vphantom{\dag}}}=\frac{\pi g^2}{\Gamma_{\rm C}}\left[AN_{1,\rm XX}+
 A'N_{2,\perp}
 \right]\:
\end{align}
where
\begin{align}\label{eq:AA}
 A=\frac{2}{\pi (3\Gamma_{\rm C}+5\Gamma_{\rm X}+P_{\rm X})},
A'=\frac{2}{\pi (3\Gamma_{\rm C}+\Gamma_{\rm X}+5P_{\rm X})}\:,\\
 G=\frac{\pi g^2 AA'}{\Gamma_{\rm C}+2\Gamma_{\rm X}+2P_{\rm X}}\:,
  N_{2\circlearrowright}=\frac{\Gamma_{\rm C} N^{(2)}_{{\rm C},x}}{\Gamma_{\rm C}+\Gamma_{\rm X}+P_{\rm X}}\:.\nonumber
\end{align}
The correlator $N_{1,\rm XX}$ is readily found from the linear system
\begin{align}
 -&(\Gamma_{\rm C}+4P_{\rm X}) N_{1G}+4\Gamma_{\rm X} N_{1G}=-2\pi g^2 N_{{\rm X},x} \varrho_1,\label{eq:long1}\\
 -&(\Gamma_{\rm C}+\Gamma_{\rm X}+P_{\rm X}) N_{1X}+N_{1G}P_{\rm X}+N_{1XX}\Gamma_{\rm X}=0,\nonumber\\
 -&(\Gamma_{\rm C}+4\Gamma_{\rm X}) N_{1XX}+4N_{1X}P_{\rm X}=0\:.\nonumber
\end{align}
Here the quantities $ N_{1,\rm G}$, $ N_{1,\rm X}$ and $N_{1,\rm XX}$  determine the probabilities to find one exciton photon and the dot in ground, excitonic and biexciton states, respectively.
The correlators 
$N_{2,\parallel}$ and 
$N_{2,\perp}$ are  found from
\begin{align}
 -&(\Gamma_{\rm C}+4P_{\rm X}) N_{2,\rm G}+\Gamma_{\rm X}(N_{2,\parallel}+N_{2,\perp}+2N_{2\rm dark})=0,\label{eq:long2}\\
-&(\Gamma_{\rm C}+\Gamma_{\rm X}+P_{\rm X}) N_{2\parallel}+N_{2,\rm G}P_{\rm X}+N_{2,\rm XX}\Gamma_{\rm X}=\nonumber\\&\hspace{4cm}-2\pi g^2 N_{\rm XX} \varrho_2,\nonumber\\
-&(\Gamma_{\rm C}+\Gamma_{\rm X}+P_{\rm X}) N_{2\perp}+N_{2, \rm G}P_{\rm X}+N_{2,\rm XX}\Gamma_{\rm X}=0,\nonumber\\
-&(\Gamma_{\rm C}+\Gamma_{\rm X}+P_{\rm X}) N_{2,d}+N_{2,\rm G}P_{\rm X}+N_{2,\rm XX}\Gamma_{\rm X}=0\nonumber\\
-&(\Gamma_{\rm C}+4\Gamma_{\rm X}) N_{2,\rm XX}+(N_{2,\parallel}+N_{2, \perp}+2N_{2,d})P_{\rm X}=0\:.\nonumber
\end{align}
Here $N_{2,\rm G}$, $N_{2,\parallel}$,
$N_{2,\perp}$, $N_{2,d}$, and $N_{2,\rm XX}$ are the probabilities
to find one $x$-polarized biexciton photon and the dot in the 
ground state, $x$ exciton state, $y$ exciton state, dark exciton state and biexciton state, respectively.

Let us comment on the calculation details.
The equations for $\rho_1$ and $\rho_3$ are similar to Eq.~\eqref{eq:interm} and yield the denominators  given by the pumping-dependent decay rates, see Eq.~\eqref{eq:AA}. Final equation to find
the two-photon density matrix $\rho^{(2)}$, corresponding to $j=4$, is trivial, since we are interested only in the  value of $\rho^{(2)}$ traced over the states of the dot. This equation yields the prefactor $\propto 1/\Gamma_{\rm C}$ in Eqs.~\eqref{eq:Npar}--\eqref{eq:Nperp}.
 However, equations \eqref{eq:long1} and \eqref{eq:long2} to find the components of the density matrix $\rho_2$ become nontrivial. 
These equations explicitly depend on the pumping. 
For instance, the quantity $N_{2,\perp}$ describes the probability, that the $y$-polarized exciton is generated in the quantum dot, after the $x$-polarized biexciton is emitted. Such process contributes to the correlator \eqref{eq:Nperp} and destroys the entanglement. It is illustrated on Fig.~\ref{fig:Cascade}b.  Another similar pumping-induced process is the the absorption of the biexciton in the dot after the emission of the exciton photon. It is described by the correlator $N_{1,\rm XX}$ and also contributes to Eq.~\eqref{eq:Nperp}\:.
At the vanishing pumping both these  processes are quenched, which leads to the totally entangled state with $N_{\parallel}=N_{\circlearrowright}$, $N_{\perp}=0$.
\end{document}